# A Design of Endurance Queue for Co-Existing Systems in Multi-Programmed Environments

Subrata Ashe

TATA Consultancy Services

subrata.ashe@tcs.com

**ABSTRACT**

These days' enterprise applications try to integrate online processing and batch jobs into a common software stack for seamless monitoring and driverless operations. Continuous integration of these systems results in choking of the poorly performing sub-systems, when the service demand and throughput are not synchronized. A poorly performing sub-system may become a serious performance bottleneck for the entire system if its serviceability and the capacity is over utilized by increased service demand from upstream systems. From all the integrated sub-systems, queuing systems are majorly categorized as choking elements due to their limited service length and lack of processing details. The situation becomes more pronounced in multiprogramming environments where the queue performance exponentially degrades with increased degree of multiprogramming at upstream levels.

This paper presents an approach to compute the queue length and devise a distribution model such that the queue length is dynamically adjusted depending on the sudden growth or decline of transmission packets. The idea is to design a heat map of the memory and correlate it with the queue length distribution. With each degree of multi-programmability, the data processing logic is adjusted by the distribution model to arrive at an endurance level queue for long term service under variable load conditions. It will take away the current implementation of using delayed processing logic and/or batch processing of data at downstream systems.

## 1. INTRODUCTION

The advent of data intensive applications has made legacy enterprise systems to co-exist for seamless operations. These applications generate large amount of processed data which transmits to downstream systems for further processing. With an exponentially increasing growth pattern, these applications stress out the downstream systems which are mainly formed from legacy systems. For increasing the end to end performance of the enterprise, the growth model has to be synchronized across the applications. However, legacy systems cannot be adjusted for catering to exponential data growth and will fail under sustained load greater than the achievable load for longer duration. As an example, in an ecommerce platform depending on a MQ services for pricing calls will not be able to handle dense loads for shorter durations and sparse load for longer duration. Also load density with respect to time also plays an important aspect in this sustainability model. All these factors are studied in this paper to design an abstract model of endurance queue such that it can act as a logic gate to throttle the density as per service demand. We use the

Peter-Ackermann function [1,4] to design the dynamic growth pattern and wrap the queue with the data density logic calculated from the convolution algorithm [2,3].

## 2. PETER-ACKERMANN FUNCTION

Ackermann function [1,4] is a total computable function that is not primitive recursive. A two argument Ackermann is known as Peter-Ackermann function [1,4] which is defined as follows for nonnegative integers *m* and *n*:

$$A(m,n) = \begin{cases} n+1 & \text{if } m=0 \\ A(m-1, 1) & \text{if } m>0, n=0 \\ A(m-1, A(m,n-1)) & \text{if } m>0, n>0 \end{cases}$$

A major advantage of using this function is because it grows very rapidly for small inputs. Each time there is a decrease in *n*, the value of m also decreases and reaches to *0* after which it takes the static form of *n+1*. However, the value of *n* is not regulated and it may increase greatly with decrease in *m*. The selection of this function for designing the endurance queue is because of the typical pattern in which it gets calculated.

## 3. ENDURANCE QUEUE

An endurance queue is a temporary data storage entity whose capacity is decided based on the throughput of upstream systems and downstream systems such that the data can be serviced in a proper way without causing substantial processing delays. These queues will not be implemented at the online processing channels, but will be servicing systems that are near-real time and batch[5]. The objective of this queue is to make the entire software stack highly available and try to utilize the system resources to its maximum. It will be designed to work for longer duration without overflowing and sustain a variable load with respect to the service time of upstream systems. The queue depth will be maintained by the Peter-Ackermann growth where maximum data density interprets to maximum queue depth. The logic is adapted to multi-programming environments where-in each growth state will be spawned across processor threads and these states will decide on the queue depth for that multiprogramming level.

## 4. DESIGN OF THE ENDURANCE QUEUE

The endurance queue is developed over the existing queue. It consists of two components:

1. Monitor algorithm for queue depth
2. Multi-program level

To design the queue system, the systems are analyzed in terms of throughput(p), utilization (u), service demand(D), data density(dt) and service time(S). The end to end throughput(P) and utilization(U) is also calculated for the entire stack.

For each system, $S_i$, the capacity $C_i$ is calculated as

$$C_i = \left(\frac{p}{P} \times \frac{u}{U}\right) + \left(\frac{dt}{D} \times S\right) \qquad (1)$$

This individual system capacity $C_i$ is compared against each other to find the maximum and minium capable systems. A disjoint set X(s) of systems are formed as below:

$$\sup X(s_i) < \sup (X(s_{i+1})) \text{ and}$$
$$X(S_i(C_i)) \leqq X(S_i(C_{i+1})) \qquad (2)$$

Now we analyse the data density over the entire duration and design a heat map based on amount of data flow and service demand for each system as per fig.1

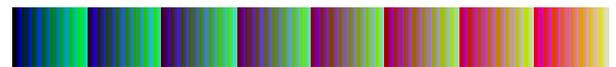

Figure 1: Heat map of systems based on data density and service demand. Highest density is shown in deep color gradient. Horizontal color lines represent peak service demand during that period mapped over the overall data density

From the heat map, the position of endurance queue is calculated to start with its first queue element. For consecutive memory blocks, a single queue can be created which will be switched for position within the memory block. A reference pointer will be carried by the sup X($s_i$) of that disjoint set so that once the set achieves $\left(\frac{u}{U}\right) = 1$, the pointer will be dereferenced from that block.

### 4.1. MONITOR ALGORITHM

After the queue is estabished, we will now design the monitor algorithm that will decide when and where to grow the queue. For the queue $Q_i$ implemented between a disjoint set X($s_i$) and X($s_{i+1}$), the data density determines the queue depth pattern. For random distribution of data and service demand, the queue depth follows the A(m,n) calculation path as below:

We will consider the current service demand as Ds and the current queue depth as Qd, then for ideal situation when Ds=0,

Queue Growth, $Q_g$ = A(Ds, Qd)

$\qquad$ = A(0, Qd)

$\qquad$ = Qd+1

(3)

When there is no message in the queue, then $Q_g$=1. This is required to prevent the queue to be destroyed because it may be needed durng a random distribution of service demand.

For a long duration distribution model, the growth behavior will be $\sum(\prod_{i=1} Qg_i)$. A dynamic queue growth for A(4,2) will be

A(4,2) = $A\big(3, A(4,1)\big)$

= $A\Big(3, \big(A(3, A(4,0))\big)\Big)$

= $A\Big(3, \big(A(3, A(3,1))\big)\Big)$

= $A\Big(3, \big(A(3, A(2, A(3,0)))\big)\Big)$

= $A\Big(3, \big(A(3, A(2, A(2,1)))\big)\Big)$

= $A\Big(3, \big(A(3, A(2, A(1, A(2,0))))\big)\Big)$

= $A\Big(3, \big(A(3, A(2, A(1, A(1,1))))\big)\Big)$ $\quad$ (4)

....

= $2^{65536} - 3$

Similary for A (2,1) :

= $A(1, A(2,0))$

= $A(1, A(1,1))$

= $A(1, A(0, A(1,0)))$ ....

With each change in the heat map of fig.1, the $Q_g$ function will increment its caluclation step. For example, when the service demand increases by 1, the function $Q_g$ executes the next step to accomodate more data. This translates to a high level algorithm as below:

*Initiate Qd=0*

*Initiate $Q_g$ = Qd+1*

*For $dt_i > dt_{i+1}$*

*Do*

$\quad$ $Q_g = A(m-1, A(m, n-1))$

$\quad$ $Q_g = Q_g + 1$

$\quad$ Check r= $\frac{C_i}{C}$ =1

$\quad$ *If True Then*

$\qquad$ *Dereference the pointer*

$\qquad$ *Move the Queue to next densty point*

*End*

*End;*

### 4.2. MULTI-PROGRAM LEVEL

The above logic will work for sequential data flows. However, in case of upstream applications spawning new threads for the same work packet, a new parameter called degree of multiprogramming $d_m$ is mixed to the queue growth. This is done using the convolution algorithm, applied over the

queue growth. The steps to perform this is given below:

1. $d_m$ is calculated/observed for all the $S_i$ in $X(s_i)$.
2. Calculate the probability of the state of $Q_g$ of each $S_i$ using
$$P(Q_{g_1}, Q_{g_2}, \ldots Q_{g_m,}) = \left(\frac{D_1^{Q_{g_1}} D_2^{Q_{g_2}} \ldots D_m^{Q_{gm}}}{G(N)}\right)$$

Where $G(N) = \sum_n \left(D_1^{Q_{g_1}} D_2^{Q_{g_2}} \ldots D_m^{Q_{gm}}\right)$
3. If $Q_{g_i} \sqsubset Q_{g_j}$, then spawn $Q_{g_j}$ into multiple threads such that $Q_{g_i} = 0$
4. Total threads spawned for $Q_{g_j}$ =
$$P(Q_{g_1,} Q_{g_2,} \ldots Q_{g_m,}) \times Q_{g_i}$$

## 5. CONCLUSION

This paper discussed on an approach of dynamic queue growth allocation on a multi-programmed environment. This could model as an alternative to replace the delayed processing logic between OLTP (online transaction processing) and batch mode. It will also cater to the proper utilization of system resources and will prevent data to be missed by downstream systems in an event of sudden transaction peak which will push more data and create a localized dense data block. If these blocks stay for longer duration in the pipe, it may lead to cascaded transaction throttle. As a temporary store, this model of endurance queue will decrease the occurance of such throttle in the system and increase the overall system throughput and lead to better performance.